\documentclass{ws-hepph05}
\usepackage{graphics}
\usepackage{bm}
\usepackage{epsfig}

\begin{document}

\newcommand{\ovl}{\overline}
\newcommand{\sT}{{\scriptscriptstyle T}}
\newcommand{\nslash}{\kern 0.2 em n\kern -0.45em /}
\newcommand{\Pslash}{\kern 0.2 em P\kern -0.56em \raisebox{0.3ex}{/}}
\newcommand{\pslash}{\kern 0.2 em p\kern -0.4em /}
\newcommand{\kslash}{\kern 0.2 em k\kern -0.45em /}
\newcommand{\Sslash}{\kern 0.2 em S\kern -0.56em \raisebox{0.3ex}{/}}
\def\adj{{\phantom{h}}}   
\newcommand{\g}{\gamma}
\newcommand{\sig}{\sigma}
\newcommand{\eps}{\epsilon}
\newcommand{\st}{{\scriptscriptstyle T}}
\newcommand{\xbj}{x_{\scriptscriptstyle B}}
\newcommand{\bpt}{\bm p}
\newcommand{\bkt}{\bm k_T}
\newcommand{\bSt}{\bm S}
\newcommand{\ba}{\begin{eqnarray}}
\newcommand{\ea}{\end{eqnarray}}
\newcommand{\beq}{\begin{equation}}
\newcommand{\eeq}{\end{equation}}
\newcommand{\slsh}[1]{\mbox{$\not\! #1$}}
\newcommand{\ph}{{\rule{0mm}{3mm}}}
\newcommand{\psibar}{\overline{\psi}}
\newcommand{\la}{\langle}
\newcommand{\ra}{\rangle}
\newcommand{\amp}[1]{\la #1 \ra}
\newcommand{\twoamp}[1]{\la \! \la \, #1 \, \ra \! \ra}
\newcommand{\half}{{1\over2}}
\newcommand{\dz}{\int \frac{d^{4}z}{(2\pi)^4}}
\newcommand{\dzp}{\frac{d^{4}z'}{(2\pi)^4}}
\newcommand{\hs}[1]{\hspace{#1}}
\newcommand{\simorder}{\raisebox{-4pt}{$\, \stackrel{\textstyle >}{\sim} \,$}}
\newcommand{\simordertwo}{\raisebox{-3pt}{$\, \stackrel{\textstyle <}{\sim} \,$}}
\newcommand{\Tr}{\text{Tr}}
\newcommand{\open}{{<\kern -0.3 em{\scriptscriptstyle )}}}

\title{\mbox{}\\[-22 mm]
Anomalous Drell-Yan asymmetry from hadronic or QCD vacuum effects 
\footnote{\uppercase{T}alk
presented at the \uppercase{I}nternational \uppercase{W}orkshop on
\uppercase{T}ransverse \uppercase{P}olarization \uppercase{P}henomena in
\uppercase{H}ard \uppercase{P}rocesses (\uppercase{T}ransversity 2005),
\uppercase{V}illa \uppercase{O}lmo, \uppercase{C}omo, \uppercase{I}taly,
\uppercase{S}eptember 7-10, 2005}
}

\author{Dani\"{e}l Boer}

\address{Dept.\ of Physics and Astronomy,\\ 
Vrije Universiteit Amsterdam,\\
De Boelelaan 1081, 1081 HV Amsterdam,\\ 
The Netherlands\\
E-mail: D.Boer@few.vu.nl}

\maketitle

\abstracts{The anomalously large $\cos(2\phi)$ asymmetry measured 
in the Drell-Yan process is discussed. Possible origins of this large 
deviation from the Lam-Tung relation are considered with emphasis on the 
comparison of two particular proposals: one that suggests it
arises from a QCD vacuum effect and one that suggests it is a hadronic 
effect. Experimental signatures distinguishing these effects are discussed.}

\section{Introduction}

Azimuthal asymmetries in the unpolarized Drell-Yan (DY) process 
differential cross
section arise only in the following way
\beq
\frac{1}{\sigma}\frac{d\sigma}{d\Omega} \propto 
\left( 1+ {\lambda} \cos^2\theta + {\mu} \sin 2\theta \, 
\cos\phi + \frac{{\nu}}{2} \sin^2 \theta \, {\cos 2\phi} \right),
\eeq
where $\phi$ is the angle between the lepton and hadron planes in
the lepton center of mass frame (see Fig.\ 3 of Ref.\cite{Boer99}).
In the parton model (order $\alpha_s^0$) quark-antiquark
annihilation yields $\lambda=1,\:\mu=\nu=0$. The leading order (LO) 
perturbative QCD corrections (order $\alpha_s^1$) 
lead to $\mu \neq 0$, $\nu \neq 0$ and $\lambda \neq 1$, such that 
the so-called Lam-Tung relation $1-\lambda -2 \nu = 0$ holds.
Beyond LO, small deviations from the Lam-Tung relation 
will arise. If one defines the quantity ${\kappa} \equiv -\frac{1}{4}
(1-\lambda-2\nu)$ as a measure of the deviation from the Lam-Tung relation,
it has been calculated\cite{Mirkes,BNM93} that at order $\alpha_s^2$ $\kappa$
is small and negative: $-\kappa \simordertwo0.01 $, for 
values of the muon pair's transverse momentum $Q_T$ of up to 3 GeV/c.

Surprisingly, the data  
is incompatible with the Lam-Tung relation and with its small 
order-$\alpha_s^2$ modification as well\cite{BNM93}. 
These data from CERN's NA10 Collaboration\cite{NA10a,NA10b} 
and Fermilab's E615 Collaboration\cite{Conway} are  
for {$\pi^- N \rightarrow \mu^+ \mu^- X$}, with $N=D$ and $W$.
The $\pi^-$-beam energies range from 140 GeV up to 286 GeV and the  
invariant mass $Q$ of the lepton pair is in the range $Q \sim 4 - 12$ GeV. 
The measured values for $\kappa$ are an order of magnitude larger than the 
order-$\alpha_s^2$ result and moreover, of opposite sign. 

Several explanations have been
put forward, but not all of them will be reviewed here.
Some unlikely explanations would be: {\it i}) 
NNLO pQCD corrections could solve
the discrepancy (but in that case the perturbative expansion itself would be 
questionable); {\it ii}) 
it could be a higher twist effect (but $Q^2 > 16$ GeV$^2$ seems too
high and according to the Fermilab data the deviation disappears at
high $x_\pi$, contrary to higher twist expectation; also, one would expect 
$\mu > \nu$, whereas in the data $\nu \gg \mu \approx 0$); 
{\it iii}) it could be a nuclear effect, since 
{$\sigma(Q_T)_W/\sigma(Q_T)_D$} is an increasing 
function of $Q_T$ (but according to Ref.\cite{NA10b} 
{$\nu(Q_T)$} shows no apparent nuclear dependence). 

The two possible explanations that will be discussed and compared here are:
{\it i}) a {QCD vacuum effect}\cite{BNM93}; 
{\it ii}) a hadronic effect, arising
from noncollinear parton configurations\cite{Boer99}.
The following will largely be based on a
recent comparative study performed in collaboration with A. Brandenburg,
O. Nachtmann and A. Utermann\cite{Boer05}.\\[-7 mm] 

\section{Explanation in terms of a QCD vacuum effect}

\begin{figure}[ht]
\begin{minipage}{12 cm} 
\begin{minipage}{5 cm} 
\centerline{\epsfxsize=1.6in\epsfbox{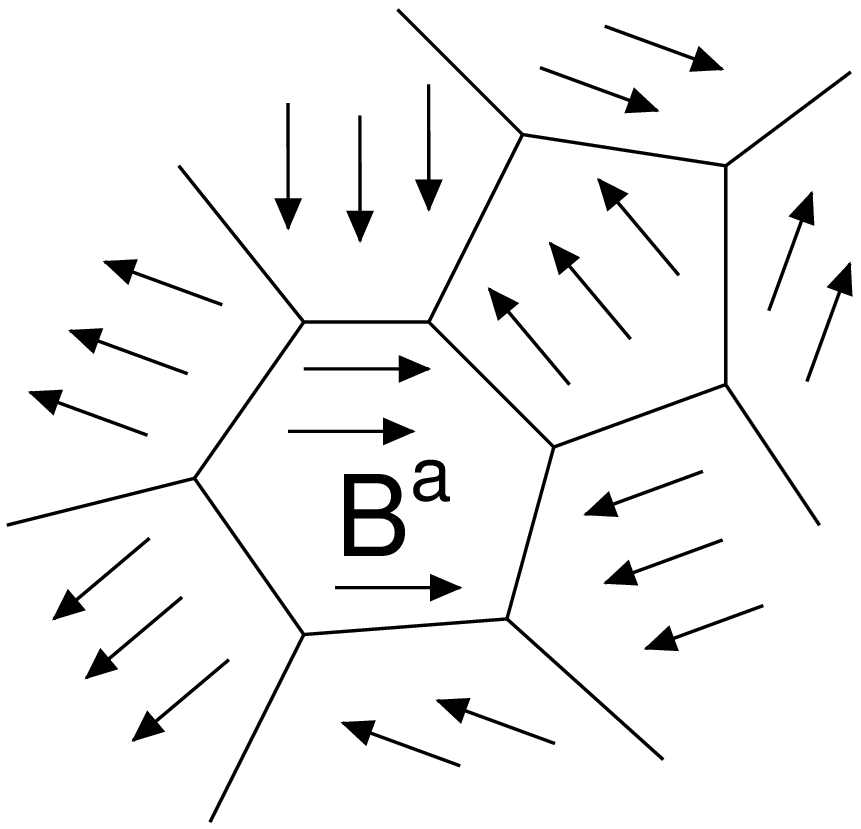}}   
\end{minipage}
\begin{minipage}{6 cm} 
\centerline{\epsfxsize=1.9in\epsfbox{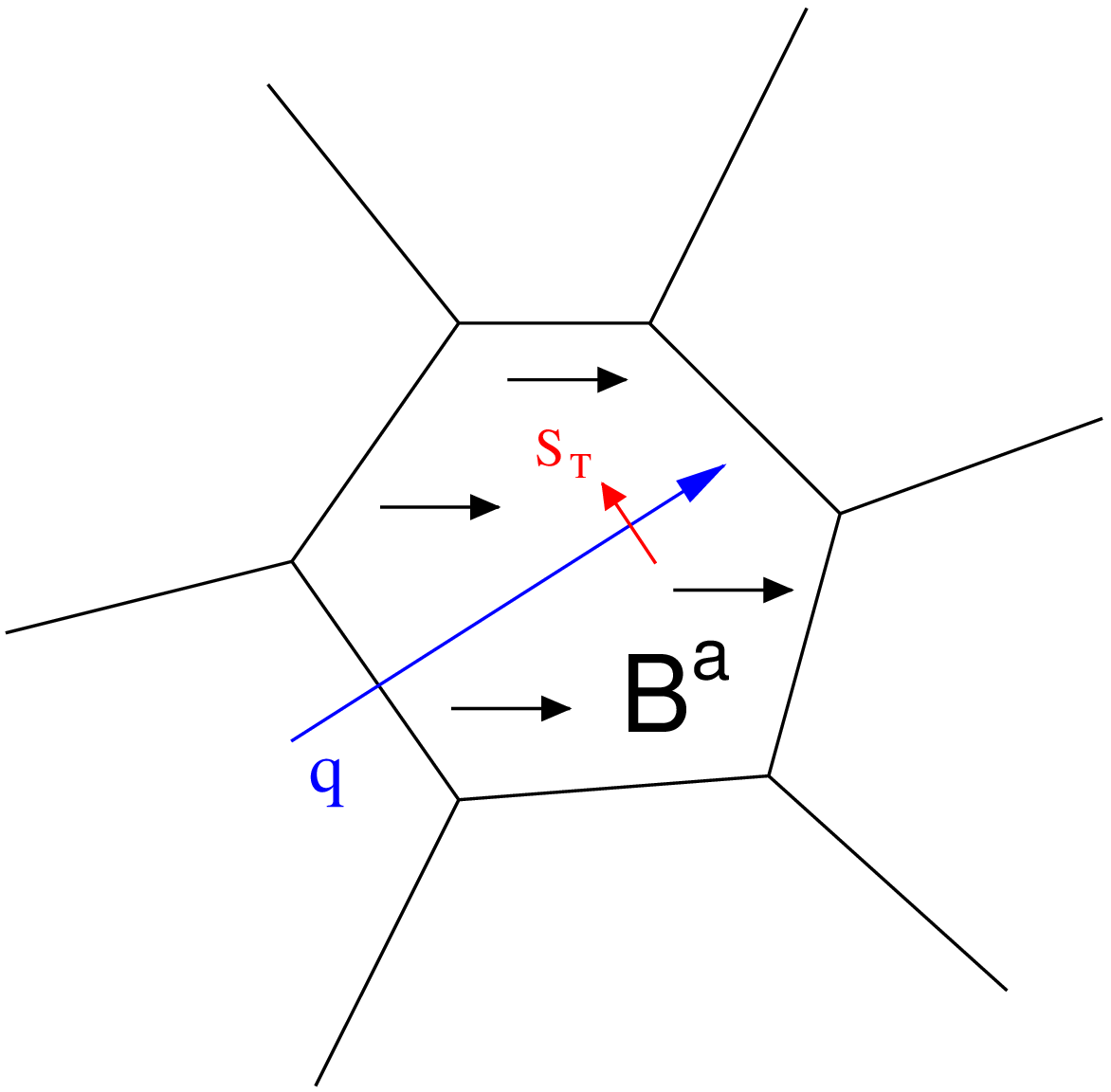}}   
\end{minipage}
\end{minipage}
\caption{Left figure: cartoon of the chromomagnetic field domain structure of
the QCD vacuum. Right figure: a fast quark traversing a domain. 
\label{domain}}
\end{figure}

Usually the DY process at $Q \sim 4 - 12$ GeV is described by
{collinear factorization}. Collinear quarks inside unpolarized hadrons are 
unpolarized themselves, implying a trivial quark-antiquark spin density matrix:
\begin{eqnarray}
{\rho^{(q,\bar{q})}}&{=}&{\frac{1}{4}\left\{\bm{1}\otimes
\bm{1}\right\}}.
\label{trivialdensity}
\end{eqnarray}
The QCD vacuum may alter this. The gluon condensate leads to
a chromomagnetic field strength (Savvidy; Shifman, Vainshtein, Zakharov; ...)
\beq
{\amp{g^2 \bm{B}^a(x) \cdot \bm{B}^a(x)} \approx (700 \, \text{MeV})^4 },
\eeq
with gluon fields having a typical 
{correlation length $a \approx 0.35$ fm} in Euclidean space. Taking this
to be an invariant length in Minkowski 
space\cite{Nachtmann-Reiter} leads to the picture of a fluctuating 
{domain structure} of the vacuum with typical domain size
$a$, schematically depicted in Fig. \ref{domain}. 
If a fast hadron, and with it a fast quark, 
traverses this domain structure, the 
time for traversing a vacuum domain is of the order of the correlation
length: {$t \approx a$}. Due to the presence of a background chromomagnetic
field the quark will acquire a transverse polarization (the 
{Sokolov-Ternov effect}). The time to build up 
{transverse polarization} is estimated\cite{Nachtmann-Reiter,bhn} to be 
much shorter than the time it takes to traverse the domain, i.e.\ 
${t \ll a}$. The radiated gluons/photons are just part of 
the cloud of virtual particles; in other words, they are included in the wave 
function. There will be no average polarization. However, if
the quark will annihilate with an antiquark in a high energy scattering 
experiment, such as DY, the polarization of the quark and the antiquark may be 
correlated if
they annihilate within a certain domain. 
\begin{figure}[ht]
\centerline{\epsfxsize=1.9in\epsfbox{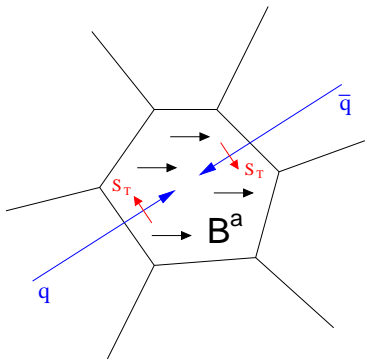}}   
\caption{Annihilation of a quark and an antiquark inside the same domain. 
\label{domainpol}}
\end{figure}
Therefore, the QCD vacuum can 
induce a {spin} {correlation} between an annihilating $q \, \bar{q}$ pair.
The quark-antiquark spin density matrix Eq.\ (\ref{trivialdensity}) 
will then be modified into 
\begin{eqnarray}
\rho^{(q,\bar{q})}&=&\frac{1}{4}\left\{\bm{1}\otimes \bm{1}+
{F_j}\,\bm{\sigma}_j\otimes \bm{1}+{G_j}\,\bm{1}\otimes 
\bm{\sigma}_j +{H_{ij}}\,
\bm{\sigma}_i\otimes \bm{\sigma}_j\right\}.
\end{eqnarray}
Only if {$H_{ij} = F_i G_j$}, then the spin density matrix {factorizes}.
But this is not necessarily so, in which case it could be called {entangled}. 
Brandenburg, Nachtmann 
\& Mirkes\cite{BNM93} demonstrated that the  
diagonal elements $H_{11}$ and $H_{22}$ can give rise to a deviation from the
Lam-Tung relation: 
\beq
{{\kappa} \equiv -\frac{1}{4} (1-\lambda-2\nu)} \approx \bigg\langle 
{\frac{H_{22}-H_{11}}{1+H_{33}}}\bigg\rangle .
\eeq
A simple assumption for the transverse momentum dependence of 
$(H_{22}-H_{11})/({1+H_{33}})$ produced a good fit to the data:
\beq
\kappa=\kappa_0\,\frac{Q_T^4}{Q_T^4+m^4_T}\, ,\hspace{1cm} \mbox{${\rm with}$}
\   
\kappa_0=0.17 \ \mbox{${\rm and}$} \ m_T=1.5~{\rm GeV}.
\eeq
Note that for this Ansatz $\kappa$ approaches a constant value ($\kappa_0 $)
for {large $Q_T$}. In other words, {the vacuum effect could persist out to 
large values of $Q_T$}. {The $Q^2$ dependence of the vacuum effect is not 
known}, but there is also no reason to assume that the spin correlation due 
to the QCD vacuum effect has to decrease with increasing $Q^2$.

\section{Explanation as a hadronic effect}

Usually if one assumes that {factorization of soft and hard energy scales} in
a hard scattering process occurs, one implicitly also assumes  
{factorization of the spin density matrix}. In the present
section this will indeed be assumed, but another common assumption will be 
dropped, namely that of {\em collinear\/} factorization.  
It will be investigated what happens if one allows for transverse momentum
dependent parton distributions (TMDs). The spin density matrix of
a noncollinear quark inside an unpolarized hadron can be nontrivial. In other
words, the transverse polarization of a {noncollinear} quark inside an
unpolarized hadron in principle can have a preferred direction and the
TMD describing that situation is called $h_1^\perp$ \cite{Boer98}. 
As pointed out in Ref.\cite{Boer99} nonzero $h_1^\perp$ leads to a deviation 
from Lam-Tung relation. It offers 
a parton model explanation of the DY data (i.e.\ with $\lambda = 1$ and $\mu
=0$): $\kappa = \frac{\nu}{2} \ \propto \
h_1^{\perp}(\pi) \, h_1^{\perp}(N)$ . In this way a good fit to data
was obtained by assuming Gaussian transverse momentum dependence. 
The reason for
this choice of transverse momentum dependence is that in order to be 
consistent with the factorization of the cross section in terms of TMDs, the 
transverse momentum of partons should not introduce another large scale. 
Therefore,
explaining the Lam-Tung relation within this framework necessarily implies
that $\kappa= \frac{\nu}{2} \to 0$ for large $Q_T$. This offers a possible
way to 
distinguish between the hadronic effect and the QCD vacuum effect. 

It may be good to mention that not only a fit of $h_1^\perp$ to data has 
been made (under certain assumptions), also several model calculations 
of $h_1^\perp$ and some of its resulting asymmetries have been 
performed\cite{GG,Boer02,LuMa04}, 
based on the recent
insight that T-odd {TMD}s like $h_1^\perp$ arise from the gauge link.  

\begin{figure}[ht]
\centerline{\epsfxsize=2.6in\epsfbox{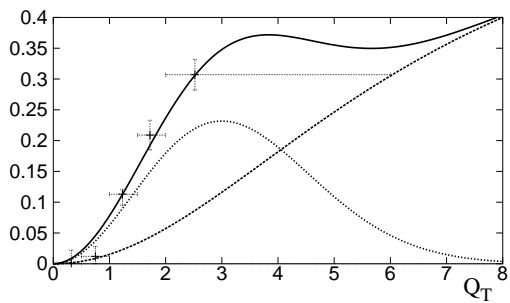}}   
\caption{{\it Impression} of possible contributions to $\nu$ as function of
$Q_T$ compared to DY data of NA10 (for $Q=8 \; \text{GeV}$). 
Dashed curve: contribution from perturbative one-gluon 
radiation. Dotted curve: contribution from a nonzero $h_1^\perp$. 
Solid curve: 
their sum.
\label{Sudpluspert}}
\end{figure}

In order to see the parton model expectation $\kappa= \frac{\nu}{2} \to 0$
at large $Q_T$ in the data, 
one has to keep in mind that the pQCD contributions (that grow
as $Q_T$ increases) will have to be subtracted. For $\kappa$ perturbative 
corrections arise
at order $\alpha_s^2$, but for $\nu$ already at order $\alpha_s$. To be
specific, at large $Q_T$ hard gluon radiation (to first order in 
$\alpha_s$) gives rise to\cite{Collins79}
\beq
{\nu(Q_T)} = \frac{Q_T^2}{Q^2+ \frac{3}{2} Q_T^2}.
\label{nuQT}
\eeq
Due to this growing large-$Q_T$ 
perturbative contribution the fall-off of the $h_1^\perp$ contribution will
not be visible directly from the behavior of $\nu$ at large $Q_T$. 
Therefore, in order to use $\nu$ as function of $Q_T$ to 
differentiate between effects, 
{it is necessary to subtract the calculable pQCD contributions}. 
In Fig.\ \ref{Sudpluspert} an illustration of this point is given. 
The dashed curve 
corresponds to the contribution of Eq.\ (\ref{nuQT}) at $Q=8 \; \text{GeV}$. 
The dotted line is a possible, parton model level, contribution from 
$h_1^\perp$ with Gaussian transverse momentum dependence. Together these
contributions yield the solid curve (although strictly speaking it is not the
case that one can simply add them, since one is a noncollinear 
parton model contribution
expected to be valid for small $Q_T$ 
and the other is an order-$\alpha_s$ result
within collinear factorization expected to be valid at large $Q_T$). 
The data are from the NA10 Collaboration 
for a pion beam energy of 194 GeV/c \cite{NA10b}. 

The $Q^2$ dependence of the $h_1^\perp$ contribution is not known to date. 
Only the 
effect of resummation of soft gluon radiation on the $h_1^\perp$ contribution 
to $\nu$ (and $\kappa$) has been studied to some extent and was found to be
quite important \cite{Boer01}. 
It gives rise to a considerable Sudakov suppression
with increasing $Q$: in going
from $Q=10$ to $90$ GeV, the contribution decreases by an order of
magnitude and approximately follows a $1/Q$ behavior 
(although it is neither a dynamical nor a kinematical higher twist effect). 
Interestingly, the contribution from 
hard gluon radiation (\ref{nuQT}) 
decreases more rapidly: as $1/Q^2$ at fixed $Q_T$.
But it seems safe to conclude that using the $Q^2$ dependence of $\nu$ 
(or $\kappa$) to differentiate between effects is not feasible at present. 

By assumption, nonzero $h_1^\perp$ gives rise to a factorized product of 
spin density
matrices ${\rho^{(q,\bar{q})}} = {\rho^{(q)}}\otimes
{\rho^{(\bar{q})}}$ with\cite{Boer05}
\begin{eqnarray}
{\rho^{(q)}}&=&\frac{1}{2}\left\{\bm{1}+\frac{h_1^\perp}{f_1}
\frac{x_1}{M_1} \left(\bm{e}_3 \times \bm{p}_{1} \right) 
\cdot \bm{\sigma} \right\} 
\ \ 
\equiv \ \ 
\frac{1}{2}\left\{\bm{1}+
F_j\,\bm{\sigma}_j\right\},\\
{\rho^{(\bar{q})}}&=&\frac{1}{2}\left\{\bm{1}-
\frac{\bar{h}{}_1^\perp}{\bar{f}{}_1}  
\frac{x_2}{M_2} \left(\bm{e}_3 \times \bm{p}_{2} \right) 
\cdot \bm{\sigma}\right\}\ \ 
\equiv \ \ 
\frac{1}{2}\left\{\bm{1}+
G_j\,\bm{\sigma}_j \right\}. 
\end{eqnarray}
Therefore, {$H_{ij} = F_i G_j$} with {$H_{33}=0$}. 
Unfortunately it is hard to observe the difference between 
$H_{33}=0$ and $H_{33} \neq 0$. 
But the factorization $H_{ij} = F_i G_j$ should shows itself via 
{consistency among various processes}, which is based on the fact that the same
function $h_1^\perp$ appears in different processes. 
Regarding this universality, complications  
have recently been addressed\cite{Bacch} that go 
beyond the sign change\cite{Collins02} that occurs 
between semi-inclusive DIS ($e\, p \to e' \, \pi \, X$) and DY: 
${(h_{1}^\perp)_{{\rm SIDIS}}} \ = \  {- (h_{1}^\perp)_{{\rm DY}}}$.
Nevertheless, the different numerical factors with which $h_1^\perp$ arises 
in different processes are calculable (functions of $N_c$ only) and 
can be taken into account. 
  
\section{Hadronic effect versus vacuum effect}

Summarizing the features of the two approaches in a table:  
\begin{table}[h]
\tbl{Comparison of the hadronic and the vacuum effect}
{\footnotesize
\begin{tabular}{@{}lll@{}}
\hline
{} &{} &{}\\[-1.5ex]
{} & $h_1^\perp \neq 0$ & QCD vacuum effect\\[1ex]
\hline
{} &{} &{}\\[-1.5ex]
$\rho^{(q,\bar{q})}$ & $\rho^{(q)}\otimes \rho^{(\bar{q})}$ & possibly
entangled\\[1ex]
{$Q$ dependence} & {$\kappa \stackrel{?}{\sim} 1/Q$} & {?} \\[1ex]
{$Q_T \to \infty$} & {$\kappa \to 0$} & need not disappear 
({$\kappa \to \kappa_0$})\\[1ex]
flavor dependence & yes & {flavor blind} \\[1ex]
$x$ dependence  & yes  
& yes, but flavor blind \\[1ex]
\hline
\end{tabular}\label{table1}}
\end{table}

\noindent
As indicated in the table, the hadronic effect will generally be flavor 
dependent and have an $x$ dependence
that is flavor dependent, since there is no reason to assume that
$h_1^\perp$ for the $u$ quark should be the same as (or simply related to) 
that for the $d$ quark. This is different from the QCD vacuum effect, 
which in this sense is flavor blind; it
does not matter whether the spin correlation is between
$u \, \bar u$ or $d \, \bar d$ (except for presumably small mass
corrections). There will be an $x$ dependence, since that determines the
energy of the annihilation process, but this again should be flavor blind. 
It should be emphasized that flavor blindness in general does not imply hadron
blindness or even process blindness. So the best next step would be 
to perform experiments with different beams ($\pi^{+}, p, \bar{p},
\ldots$, where $\pi^{+}$ and $\bar{p}$ offer the advantage of having 
valence anti-quarks) and in different kinematical regimes. 
For instance, the measurement of $\langle\cos 2\phi \rangle$ can be done at 
{\small RHIC} in $p \, p \rightarrow \mu^+ \mu^- X$, or in 
$p \, \bar p \rightarrow \mu^+ \mu^- X$ at Fermilab or {\small GSI/FAIR}. 

The use of polarized beams can also help 
(e.g.\ at {\small RHIC} or {\small GSI}). In the DY process with one
transversely polarized hadron, the differential cross section can 
namely depend on 
the azimuthal angle $\phi_S$ of the transverse hadron spin (${\bm S}_T$)
compared to the lepton plane: 
\[
\frac{d\sigma(p\, p^\uparrow \to \ell \, \bar \ell \, X)}{d\Omega\; 
d\phi_{S}} \propto 
1+ \cos^2\theta 
+ \sin^2 \theta \left[ \frac{{\nu}}{2} \; \cos 2\phi - {\rho} \; 
|\bm S_{T}^{}|\;
{\sin(\phi+\phi_{S})} \right] + \ldots 
\]
Within the framework of TMDs the analyzing power $\rho$ is 
proportional to the product $h_1^\perp \, h_1 $ \cite{Boer99}, which involves 
the transversity function $h_1$. A nonzero function $h_1^\perp$ will provide 
a relation between $\nu$ and $\rho$,
which in case of one (dominant) flavor (usually called $u$-quark 
dominance) and Gaussian transverse momentum dependences, 
is approximately given by 
\beq
\rho \approx
\frac{1}{2} \, 
\sqrt{\frac{{\nu} }{{\nu_{\mbox{{\small ${\rm max}$}}}}}}\, \frac{h_1}{f_1},
\label{approx}
\eeq
where $\nu_{\mbox{{\small ${\rm max}$}}}$ is the maximum value attained by 
$\nu (Q_T)$.
This relation depends on the magnitude of $h_1$ compared to $f_1$ (see 
Refs.\cite{Boer99b,Boer02b} for explicit examples) and 
this may be extracted from double transverse spin asymmetries in DY
(potentially at {\small RHIC} or {\small GSI}) or from 
SIDIS data (from e.g.\ {\small HERMES} or {\small COMPASS}) 
by exploiting the interference fragmentation
functions (which can be obtained from $e^+ e^-$ data, e.g.\ at {\small 
BELLE}). 

Also semi-inclusive DIS can be used. The $\left< {\cos
2\phi} \right>$ in {$e\, p \to e' \, \pi \, X$} 
would be {$\propto h_1^\perp H_1^\perp$}, where $H_1^\perp$ is the Collins
fragmentation function (also obtainable from {\small 
BELLE}). This particular SIDIS observable has been studied using model 
calculations\cite{Gambergetal}. All this illustrates how the consistency 
among processes may be used to test the $h_1^\perp$ hypothesis. 

\section{Conclusions}

A transverse spin correlation in quark-antiquark annihilation ($q^\uparrow 
\bar{q}^\uparrow \to \gamma^*$) will lead to a
{$\cos(2\phi)$ asymmetry} in the DY lepton-pair angular distribution. 
Such a spin correlation can arise from the chromomagnetic background field in
the {QCD vacuum} or from {noncollinear partons}. If a 
flavor dependence is observed
in future data, it would favor a hadronic effect. On the other hand, 
persistence of the asymmetry at 
large values of $Q_T$ and $Q$ (after subtraction
of pQCD corrections if needed) would favor a vacuum effect. 
Several future and ongoing experiments will be able to 
provide crucial information on these dependences. 

\section*{Acknowledgments}
I thank Arnd Brandenburg, Stan Brodsky, Dae Sung Hwang, Otto Nachtmann and 
Andre Utermann for fruitful discussions and collaboration on this topic. 
The research of D.B.~has been made possible by 
financial support from the Royal Netherlands Academy of Arts and Sciences.

\end{document}